\documentclass[runningheads]{llncs}
\usepackage[T1]{fontenc}
\usepackage{graphicx}
\usepackage{hyperref}
\usepackage{amsmath}
\usepackage{amssymb}
\usepackage{subcaption}
\usepackage{booktabs}
\usepackage{comment}
\usepackage{multirow}
\usepackage[table]{xcolor}
\usepackage{array}
\usepackage{soul}
\usepackage{xcolor}
\usepackage{tikz}
\usepackage{cite}
\begin{document}
\title{Between Zeros and Ones: Behavioral Characterization Beyond Binary Labeling Across Public ICS Datasets}
%
%
\author{Konstantinos E. Kampourakis\inst{1}\orcidID{0009-0000-8883-0735} \and
Vyron Kampourakis\inst{1}\orcidID{0000-0003-4492-5104} \and
Georgios Spathoulas\inst{1}\orcidID{0000-0003-2947-486X} \and
Constantinos Kolias\inst{2}\orcidID{0000-0002-3020-291X}}
\authorrunning{V. Kampourakis et al.}
%
\institute{Norwegian University of Science and Technology, 2802 Gjøvik, Norway \email{\{konstantinos.kampourakis, vyron.kampourakis, georgios.spathoulas\}@ntnu.no} \and  University of Idaho, Idaho Falls, ID 83402, USA \email{kolias@uidaho.edu}}

\maketitle              
\begin{abstract}

Intrusion detection in Industrial Control Systems (ICS) is typically evaluated on a small set of public benchmarks using binary ``normal'' versus ``attack'' labels, a practice that can mask the behavioral diversity of cyber-physical attacks. To address this limitation, we propose a behavioral characterization framework that maps raw multivariate process traces into five interpretable physical primitives: drift, spike, oscillation, repetition, and switching. We apply the framework to three widely used ICS benchmarks, namely, SWaT, WADI, and HAI, and show that attack windows exhibit clear behavioral shifts relative to normal operation while the three datasets occupy largely distinct regions of the behavioral space, revealing both cross-dataset bias and intra-dataset diversity. In particular, WADI is dominated by repetition, HAI emphasizes sustained drift and oscillation, and SWaT is characterized by stealthier frozen-telemetry behavior. To examine the evaluation implications, we use an indicative Random Forest baseline and show that aggregate binary metrics can limit visibility into performance across different behavioral proxies. For example, in SWaT, macro F1 drops from 85.44\% under binary evaluation to 37.84\% under behavior-proxy multiclass prediction, with similar degradations observed on WADI and HAI. Based on these findings, we argue for complementing conventional binary benchmarking with behavior-stratified evaluation to expose blind spots that aggregate scores leave hidden and to better support targeted incident response.

\end{abstract}
\section{Introduction}
\label{S:intro}

Public ICS datasets such as SWaT~\cite{mathur2016}, WADI~\cite{ahmed2017wadi}, and HAI~\cite{Hyeok-Ki2020} have become de facto benchmarks for evaluating Intrusion Detection Systems (IDS) targeting cyber-physical attacks. Most studies treat these datasets as labeled time series in which each sample receives a binary attack-or-normal label~\cite{mustafa2025,conti2021,ahmed2025hybrid,kadosh2020}, and evaluate IDS models using aggregate metrics such as precision, recall, and F1-score. This binary labeling practice implicitly assumes that the attack class is homogeneous. In reality, industrial processes can be manipulated in many ways, including gradual sensor drifts, abrupt spikes, frozen telemetry, oscillatory control behavior, and forced actuator switching~\cite{Shen2024,Oyama2021,Zambrano2021,info17030286}. These behaviors are visible in the temporal structure of process traces but are typically not distinguished in IDS evaluation.

However, ignoring behavioral diversity can have practical consequences. Unlike traditional IT security, where the response to an intrusion is largely uniform, ICS environments require differentiated operator responses. For example, a spike indicating a valve slam may require an immediate safety trip, whereas a slow sensor drift may call for recalibration over several hours. In this context, binary detection collapses these qualitatively different urgencies into a single signal, leaving operators without the behavioral specificity needed to respond appropriately. To address this, we introduce a behavioral characterization framework that maps multivariate ICS traces into five interpretable temporal primitives: drift, spike, oscillation, repetition, and switching. 

The framework is purely observational and exploratory; namely, it quantifies how attacks may manifest in normalized process traces, but does not infer attacker intent, semantic attack types, or a universal attack taxonomy. This scope is intentional; rather than requiring process-specific domain knowledge or behavioral ground truth, the framework characterizes what is directly observable and reproducible across datasets. This makes it a portable first step toward exposing the behavioral diversity that binary labels may leave unmeasured. We apply it to SWaT, WADI, and HAI and address four research questions:

\begin{itemize}
  \item \textbf{RQ1:} Do attack windows exhibit behavioral characteristics different from normal operation?
  \item \textbf{RQ2:} Do the datasets differ in the distribution of normalized behavioral morphologies captured by our primitives?
  \item \textbf{RQ3:} Are attack events internally diverse in their behavioral structure?
  \item \textbf{RQ4:} How does binary IDS evaluation compare with proxy-based evaluation in distinguishing attack-window behaviors?
\end{itemize}

The rest of the paper is structured as follows. Section~\ref{S:rel:work} reviews related work. Section~\ref{sec:methodology} presents the framework. Section~\ref{sec:results} reports empirical findings. Sections~\ref{S:takeaways} and~\ref{S:limitations} discuss takeaways and limitations. The last section concludes and provides pointers for future work.

\section{Related Work}
\label{S:rel:work}

This section briefly discusses related work on ICS security datasets, intrusion-detection evaluation practices, and benchmark limitations relevant to our study.

Public ICS benchmarks such as SWaT~\cite{mathur2016}, WADI~\cite{ahmed2017wadi}, and HAI~\cite{Hyeok-Ki2020} are widely used for data-driven IDS research. Although additional cyber-physical security datasets have also been proposed~\cite{Taormina2018,Lemay2016,Grammatikis2020}, evaluation remains concentrated on a small set of public benchmarks~\cite{lamberts2023sok,conti2021}. Existing comparisons usually organize datasets by domain, architecture, or protocol rather than by the behavioral characteristics of the process traces they contain~\cite{lamberts2023sok}. In this work, we focus on SWaT, WADI, and HAI because they are public, process-centric, and broadly adopted benchmarks for multivariate time-series anomaly detection~\cite{ahmed2025hybrid,Jaradat2024}.

Intrusion detection in ICS has been studied using physics-based methods, ML models, and more recently, LLM-assisted schemes~\cite{urbina2016survey,Shen2024,hu2018survey,mlics2022survey,Lian2025,Adjewa2025}. Despite the methodological differences, evaluation is commonly formulated as binary classification with aggregate metrics such as precision, recall, and F1-score~\cite{Jaradat2024,ahmed2025hybrid}. This setup implicitly treats attacks as a single class, even though cyber-physical attacks may induce qualitatively different process effects, including gradual drift, abrupt transients, oscillatory behavior, stale telemetry, or excessive actuator switching~\cite{Krotofil2014,Oyama2021,Zambrano2021}. Our work departs from this detector-centric perspective by characterizing attack windows through interpretable behavioral properties and using them to stratify evaluation. 

Prior studies have already noted that ICS benchmarks capture only specific testbeds and attack scenarios; therefore, supporting limited generalization~\cite{mathur2016,Taormina2018,Hyeok-Ki2020}. Systematization of Knowledge (SoK) papers and surveys further show that IDS evaluations are often biased by dataset choice and by overreliance on aggregate metrics~\cite{lamberts2023sok,hu2018survey}. However, existing analyses do not formalize process-behavioral bias, that is, which temporal attack behaviors are over- or under-represented in standard benchmarks. To address this gap, we quantify behavioral coverage using five interpretable temporal properties and examine how SWaT, WADI, and HAI populate this space differently.

\section{Methodology}
\label{sec:methodology}

\begin{figure}[htb]
    \centering
    \includegraphics[width=\linewidth]{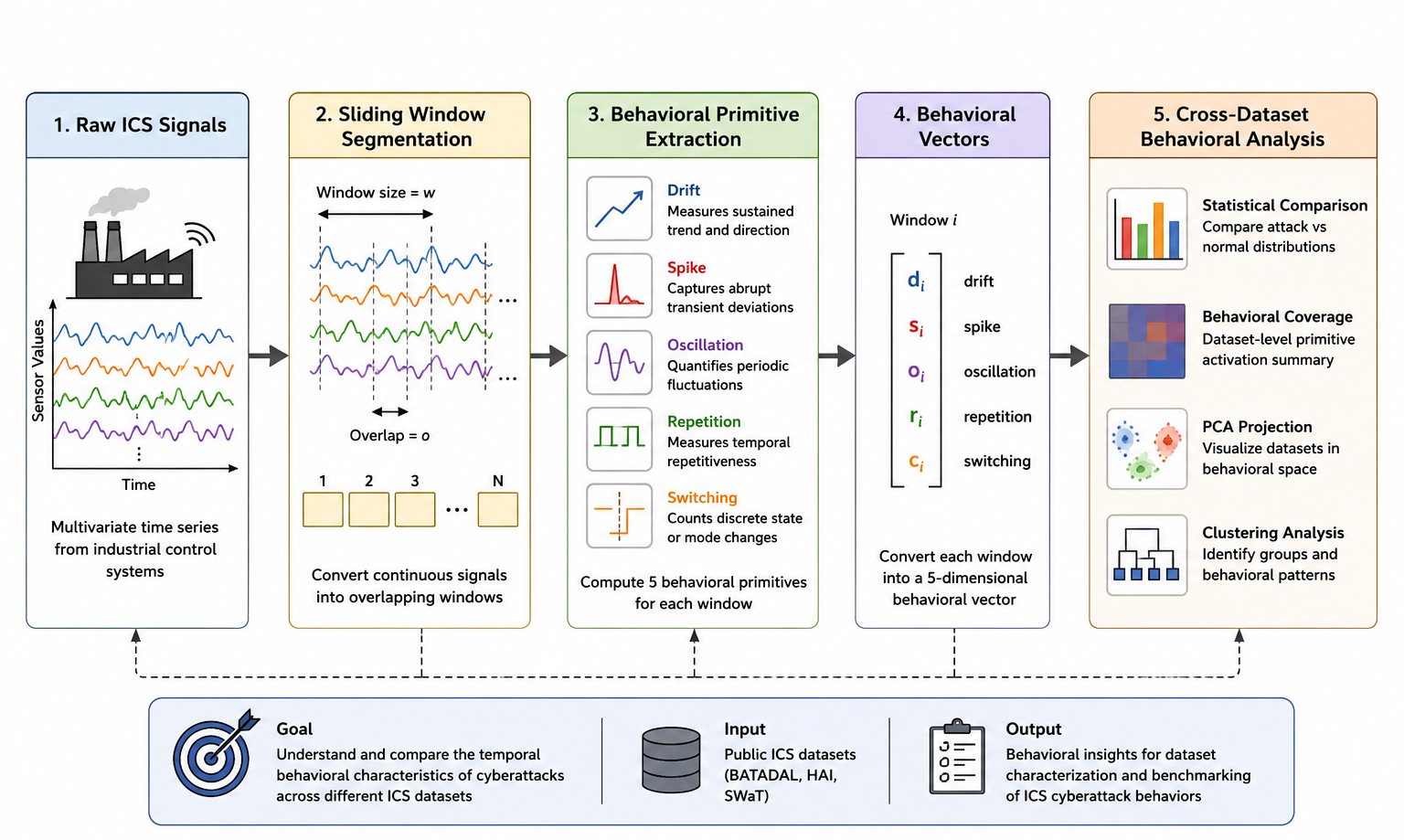}
    \caption{Overview of the proposed behavioral characterization framework.}
    \label{fig:methodology}
\end{figure}

The proposed methodology characterizes cyber-physical attacks through their observable temporal effects on industrial process dynamics, without, however, modeling attacker intent or process-specific semantics. Specifically, the framework transforms raw multivariate ICS traces into behavioral embeddings that capture generic operational disturbances across heterogeneous industrial environments. Figure~\ref{fig:methodology} provides an overview of the framework's pipeline, in which raw process traces are segmented into overlapping temporal windows, from which lightweight behavioral primitives are extracted and aggregated into low-dimensional behavioral vectors. Overall, the framework consists of four stages: preprocessing, sliding-window segmentation, behavioral primitive extraction, and behavioral embedding generation.

\subsection{Preprocessing and Windowing}
\label{SS:preprocess}

The utilized ICS datasets, i.e., SWaT, WADI, and HAI, consist of multivariate temporal traces containing both continuous process variables (e.g., tank levels, flow rates, pressures) and discrete operational states (e.g., pump activity, valve positions). Let the process state at time step $t$ consist of $N$ sensor and actuator variables. To ensure that variables with substantially different physical units contribute equitably to the behavioral analysis, all continuous process variables are normalized using z-score normalization, as seen in equation~\ref{eq:zscore_normalization}. Note that we choose z-normalization because it is simple, widely used, and provides a common scale that preserves relative deviations within each variable. The normalization statistics are computed from normal operating data to prevent attack periods from contaminating the baseline process distribution. For a continuous variable $x$ (e.g., a sensor time series), normalization is performed as:

\begin{equation}
z = \frac{x - \mu_{\text{normal}}}{\sigma_{\text{normal}} + \epsilon}
\label{eq:zscore_normalization}
\end{equation}

where $\mu_{\text{normal}}$ and $\sigma_{\text{normal}}$ denote the mean and standard deviation estimated from normal operation only, and $\epsilon$ is a small constant added for numerical stability. The latter is added in the denominator to avoid division by zero for variables with near-zero variance, without materially affecting the resulting z-scores. Discrete actuator states are intentionally left unnormalized to preserve their categorical operational semantics. Following preprocessing, the multivariate traces are segmented into overlapping temporal windows.

Specifically, the proposed framework utilizes a fixed window length of $w = 60$ samples with a 50\% overlap between adjacent windows. Regarding the latter, the 50\% overlap was used to reduce the effects of boundary fragmentation and preserve the temporal continuity between adjacent windows. Simply put, without overlap, short-duration or transitional behaviors may be split across neighboring windows, weakening their representation. On the other hand, larger overlaps may increase computational cost and redundancy, while providing limited additional temporal information. 

Therefore, the empirically selected overlap represents a compromise between temporal coverage and computational efficiency. To further justify the selection, specifically for the window length, a sensitivity analysis is provided in Appendix~\ref{app:sensitivity}. Utterly, each window receives a binary label based on the proportion of attack-labeled samples contained within the window. A window is assigned an attack label ($y_w = 1$) if at least 50\% of its constituent samples correspond to attack periods in the ground truth of the dataset. Otherwise, the window is labeled as normal ($y_w = 0$). The readers should keep in mind that the threshold selection is a practical choice rather than a universal solution; different window sizes and overlap ratios could be used, depending on the temporal resolution and dynamics of the dataset.

\subsection{Behavioral Primitive Design}
\label{SS:behPriDe}

The proposed framework employs a set of interpretable behavioral primitives derived from classical time-series statistics and does not rely on learned latent representations or process-specific feature engineering. The primitives are intentionally lightweight and computationally inexpensive, enabling applicability across heterogeneous ICS datasets without assumptions regarding process physics or control architecture. In other words, based on common empirical observations during adversarial activity in CPS~\cite{Krotofil2014,Zambrano2021,urbina2016survey}, primitives were selected according to three criteria: (i) interpretability, (ii) computational simplicity, and (iii) cross-dataset applicability. Of course, the selected primitives should be viewed as a conservative baseline rather than an exhaustive characterization library. Alternative descriptors, including entropy-based measures, spectral features, wavelets, lag-response statistics, and nonlinear dynamical metrics, are plausible candidates for future primitive-selection studies.

\begin{table}[!htbp]
\centering
\caption{Behavioral interpretation of the proposed primitives.}
\label{tab:primitive_rationale}
\resizebox{\textwidth}{!}{%
\begin{tabular}{llp{7cm}}
\toprule
\textbf{Primitive} & \textbf{Captured Behavior} & \textbf{Physical Manifestation} \\
\midrule
Drift & Sustained monotonic deviation & Sensor bias injection, gradual process manipulation, slow tank level deviation \\
Spike & Abrupt transient change & Sudden valve closure/opening, pressure surge, transient disturbance \\
Oscillation & Repeated directional reversals & Unstable control loops, actuator hunting, feedback instability \\
Repetition & Recurring temporal patterns & Replay attacks, periodic reuse of process trajectories \\
Switching & Excessive discrete state transitions & Forced actuator toggling, abnormal valve or pump switching \\
\bottomrule
\end{tabular}}
\end{table}

Table~\ref{tab:primitive_rationale} summarizes the operational interpretation of the behavioral primitives proposed. The selected primitives are chosen as a compact baseline that spans five broad and interpretable disturbance families commonly observable in process traces: sustained deviation (drift), abrupt transient change (spike), directional instability (oscillation), recurring temporal structure (repetition), and excessive discrete switching (switching), as seen in the table.

\subsection{Behavioral Primitive Extraction}

For each temporal window and for each normalized continuous signal $x$, the behavioral primitives introduced in section~\ref{SS:behPriDe} are computed. For notational simplicity, the first-order temporal difference is denoted as $\Delta x_t = x_t - x_{t-1}$. Importantly, the formulas rely only on elementary time-series operations, including variance, linear slope, maximum derivative, sign changes, and autocorrelation, in line with standard anomaly detection practice to provide strong baselines~\cite{wenig2022tsad,deepSurvey2022,kampourakis2026systematic} with simple statistical features.

\paragraph*{\textbf{Drift Intensity ($D$).}} Drift intensity captures sustained monotonic deviations and gradual process shifts. To quantify persistent trends, a linear model $x_t = mt + b$ is fitted within the window. The drift score combines trend magnitude with regression fit quality, as shown in equation~\ref{eq:drift_score}. Note that $m$ denotes the fitted slope and $R^2$ denotes the coefficient of determination. Higher values indicate stronger, persistent directional behavior.

\begin{equation}
D = |m| \cdot R^2
\label{eq:drift_score}
\end{equation}

\paragraph*{\textbf{Spike Intensity ($S$).}} Spike intensity captures abrupt transient disturbances and sudden process deviations. This is approximated using the maximum absolute first-order temporal difference, as shown in equation~\ref{eq:spike_score}. Higher values indicate sharper transient process changes.

\begin{equation}
S = \max_t |\Delta x_t|
\label{eq:spike_score}
\end{equation}

\paragraph*{\textbf{Oscillation Intensity ($O$).}} Oscillation intensity captures unstable control behavior and repeated directional reversals in process dynamics. The metric is computed as the normalized frequency of sign changes in the first derivative, as shown in equation~\ref{eq:oscillation_score}. Higher values indicate stronger oscillatory behavior.

\begin{equation}
O = \frac{1}{w} \sum_{t=2}^{w} \mathbf{1}\big[\text{sgn}(\Delta x_t) \neq \text{sgn}(\Delta x_{t-1})\big]
\label{eq:oscillation_score}
\end{equation}

\paragraph*{\textbf{Repetition Intensity ($R$).}} Repetition intensity captures recurring temporal structures and replay-like signal behavior. The metric is computed using the autocorrelation function $\rho(\ell)$ and extracting the maximum correlation over a bounded lag interval, as shown in equation~\ref{eq:repetition_score}. The restricted lag interval avoids trivial zero-lag correlations while focusing on short-term recurring temporal structure.

\begin{equation}
R = \max_{\ell \in [5,30]} \rho(\ell)
\label{eq:repetition_score}
\end{equation}

\paragraph*{\textbf{Switching Intensity ($A$).}} For discrete actuator signals $u_t \in \{0,1,\dots,k\}$, continuous statistical metrics are physically inappropriate. Instead, switching intensity quantifies actuator toggling and discrete operational state transitions by counting the number of state changes within the window, as shown in equation~\ref{eq:switching_score}. Higher values indicate more frequent operational switching activity.

\begin{equation}
A = \sum_{t=2}^{w} \mathbf{1}[u_t \neq u_{t-1}]
\label{eq:switching_score}
\end{equation}

Importantly, the afore-described definitions of the primitives should be understood as operational proxies for observable temporal behavior rather than complete physical definitions. For example, the drift metric captures sustained monotonic deviation but not all nonlinear drift forms; the spike metric captures abrupt transients but may not fully distinguish a sharp step followed by a plateau; the oscillation metric captures repeated directional reversals but not every limit-cycle regime; and the repetition metric captures short-lag recurrence but may fail to represent all replay-like or phase-shifted behaviors. Likewise, repetition should not be interpreted as a standalone anomaly signal, since industrial processes are often intrinsically repetitive during normal operation; its value lies primarily in combination with the other primitives when characterizing how attack windows differ behaviorally from nominal process progression.

\subsection{Feature Vector Construction}
\label{SS:percentile}

Cyber-physical attacks typically target only a small subset of process variables. Because of this, simply averaging behavioral scores across all variables would eliminate the localized effects of an attack. Conversely, taking the absolute maximum score across all variables could render the framework sensitive to isolated noise spikes. To strike a balance, we aggregate the primitive scores (drift, spike, oscillation, repetition, and switching) across all $N$ process variables using the 95th percentile operator. Conceptually, this means sorting all datasets' variables by how strongly they exhibit a given behavior, discarding the top 5\% to filter out isolated sensor glitches, and retaining the highest remaining value. Mathematically, this corresponds to the threshold below which 95\% of the variable scores fall, as shown in equation~\ref{eq:percentile}. Recall that the percentile is computed spatially (across the $N$ variables) and not temporally (across the entire dataset). This means that the embedding is generated on a per-window basis and enables characterization even in an online, real-time IDS evaluation.

\begin{equation}
P_{95} = Q(0.95),
\label{eq:percentile}
\end{equation}

Note that $Q(\cdot)$ denotes the empirical quantile function, which calculates percentiles directly from the datasets. This approach highlights localized attack behaviors while filtering out extreme, single-variable outliers. For each temporal window, we apply this aggregation to all five primitives to construct a compact, five-dimensional feature vector, as shown in equation~\ref{eq:aggregate}. This vector serves as the foundational representation for all subsequent statistical analyses, including the behavioral coverage evaluations in sections~\ref{SS:rq2} and~\ref{SS:rq3}.

\begin{equation}
v_w = [D_{95}, S_{95}, O_{95}, R_{95}, A_{95}] \in \mathbb{R}^5.
\label{eq:aggregate}
\end{equation}

\subsection{Dominant Behavioral Property Assignment}

For coarse behavior-stratified analysis and visualization, we assign each temporal window a single dominant behavioral property. Note that this label is a heuristic summary, not a claim that the window contains only one behavior. In practice, multiple primitives may co-occur within the same window, and some primitives may exhibit low baseline variance. Therefore, the dominant label should not be interpreted as a unique or exhaustive behavioral class. As explained in section~\ref{SS:percentile}, for each window, the behavioral embedding vector $v_w = [D_{95}, S_{95}, O_{95}, R_{95}, A_{95}]$ collects the 95th-percentile activations of the five primitives across all process variables. Recall from section~\ref{SS:preprocess} that before dominant-property assignment, the five primitive scores are normalized using z-score normalization (equation~\ref{eq:zscore_normalization}) over the normal operating data of each dataset. The resulting normalized values are then compared to determine the dominant primitive.

If the maximum normalized primitive activation falls below a threshold $\tau = 0.30$ (see Appendix~\ref{app:sensitivity}), the window is labeled as unassigned, indicating the absence of a clearly dominant behavioral signature. Otherwise, the primitive corresponding to $\max(v_w)$ is assigned as the dominant property. The threshold is not intended to represent a universal decision boundary; instead, it filters out weak activations arising from ambient process noise or low-amplitude fluctuations. For transparency, note that a limitation of this approach is that collapsing each window to a single dominant label may obscure simultaneous behavior activations and does not preserve the full richness of the underlying process trace.

\subsection{Behavioral Annotation Validation}

The evaluated datasets provide only binary attack labels (attack vs.\ normal) and no behavior-level annotations; therefore, the proposed primitive assignments cannot be validated in a standard supervised manner against ground-truth behavioral classes. In this respect, to assess whether the automated labels correspond to recognizable patterns in the traces, we performed a human sanity check on 60 representative attack windows, sampled as 20 from SWaT, 20 from HAI, and 20 from WADI. Specifically, two annotators, namely the first two authors, visually inspected the multivariate trajectories and assigned each window a dominant observable behavior: drift, spike, oscillation, repetition, switching, or unassigned. Be aware that the annotations were made independently and without access to the automated framework outputs.

\begin{equation}
\kappa
=
\frac{p_o-p_e}
{1-p_e}
\label{eq:kappa}
\end{equation}

\begin{figure*}[!htbp]
\centering
\includegraphics[width=\linewidth]{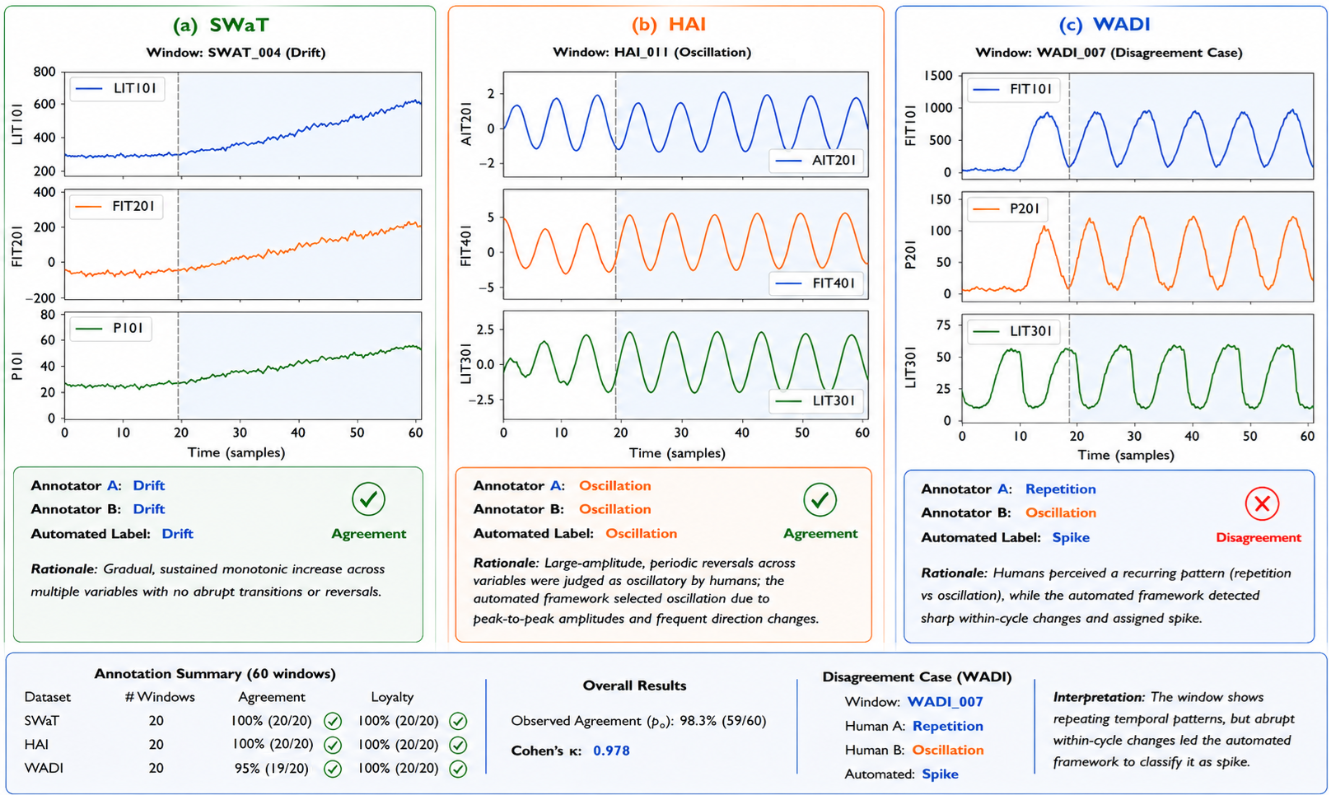}
\caption{Representative examples used during manual behavioral annotation.}
\label{F:annotator_examples}
\end{figure*}

Agreement between the framework's automated dominant label and the human annotation was measured using Cohen's $\kappa$ in Eq.~\ref{eq:kappa}. The value $p_o$ denotes observed agreement and $p_e$ the chance agreement. The comparison yielded 98.3\% observed agreement and a $\kappa$ value of 0.978, which indicates high consistency between the automated assignments and the human annotations. Only one disagreement occurred among the 60 windows, in a WADI instance where one human annotator identified repetitive activity, the other identified an oscillatory pattern, and the framework assigned a spike label. This disagreement reflects the ambiguity that can arise when recurring patterns contain both oscillatory behavior and abrupt transient fluctuations. Because the datasets provide no behavior-level ground truth and the annotators were also the framework’s designers, the sanity check should be interpreted as a limited rather than an independent validation. Nevertheless, in general terms, the results suggest that the automated primitive assignments correspond to recognizable temporal patterns in the underlying process traces. Figure~\ref{F:annotator_examples} presents representative examples from SWaT, HAI, and WADI, including the single disagreement case.

\section{Results}
\label{sec:results}

We evaluate the behavioral characterization framework proposed across the SWaT, WADI, and HAI datasets to answer the four research questions introduced in Section~\ref{S:intro}. Collectively, the results demonstrate that ICS attack traces exhibit substantial behavioral heterogeneity both across and within datasets. That is, attack windows occupy distinct regions of the proposed behavioral space depending on the temporal process dynamics induced during anomalous operation, and do not form a single homogeneous anomaly class.

\subsection{RQ1: Behavioral Shift Between Normal and Attack States}

\begin{figure*}[!htbp]
    \centering
    
    \begin{subfigure}[b]{0.48\linewidth}
        \centering
        \includegraphics[width=\linewidth]{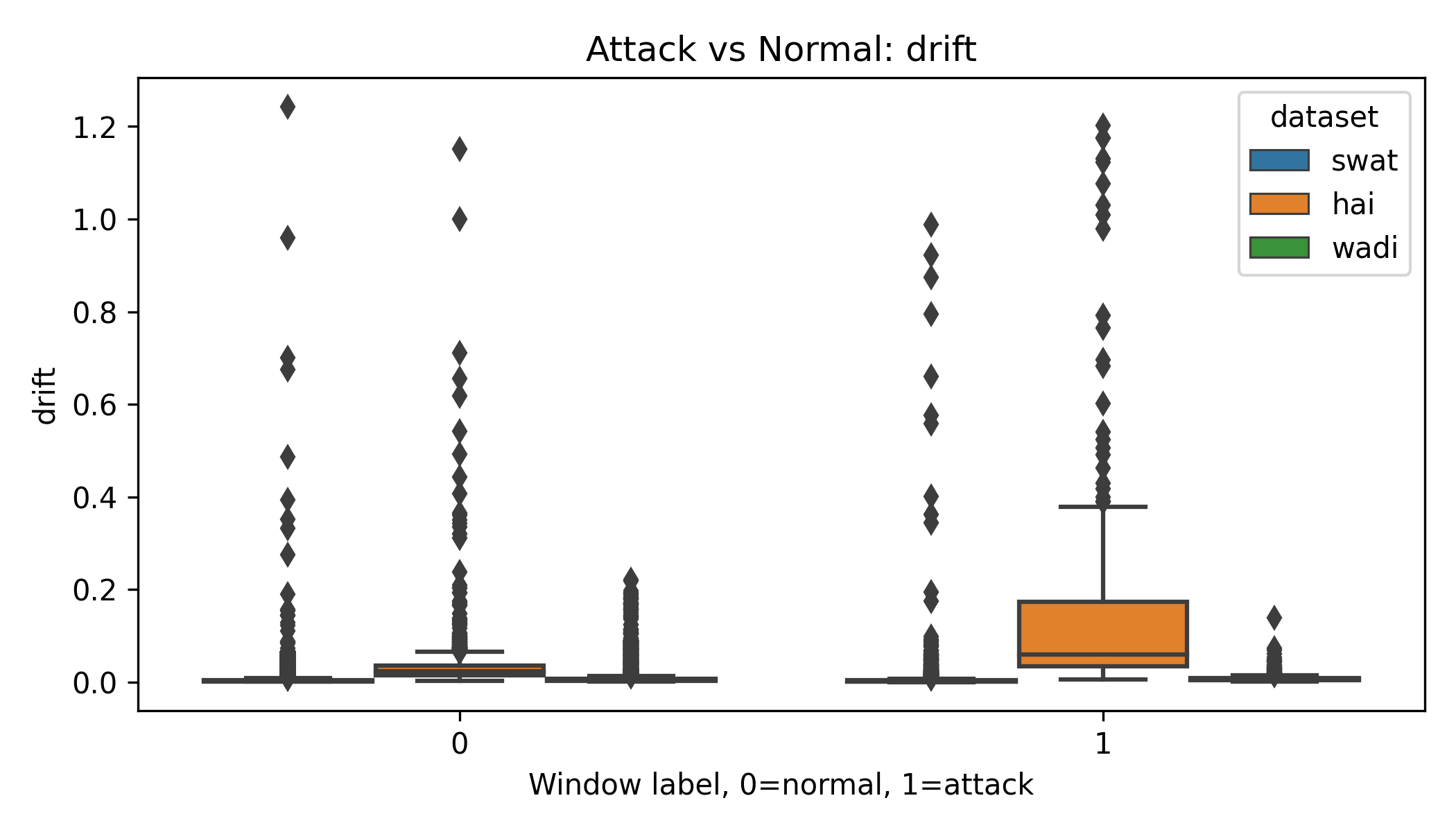}
        \caption{Drift intensity}
        \label{fig:exp1_drift}
    \end{subfigure}
    \hfill
    \begin{subfigure}[b]{0.48\linewidth}
        \centering
        \includegraphics[width=\linewidth]{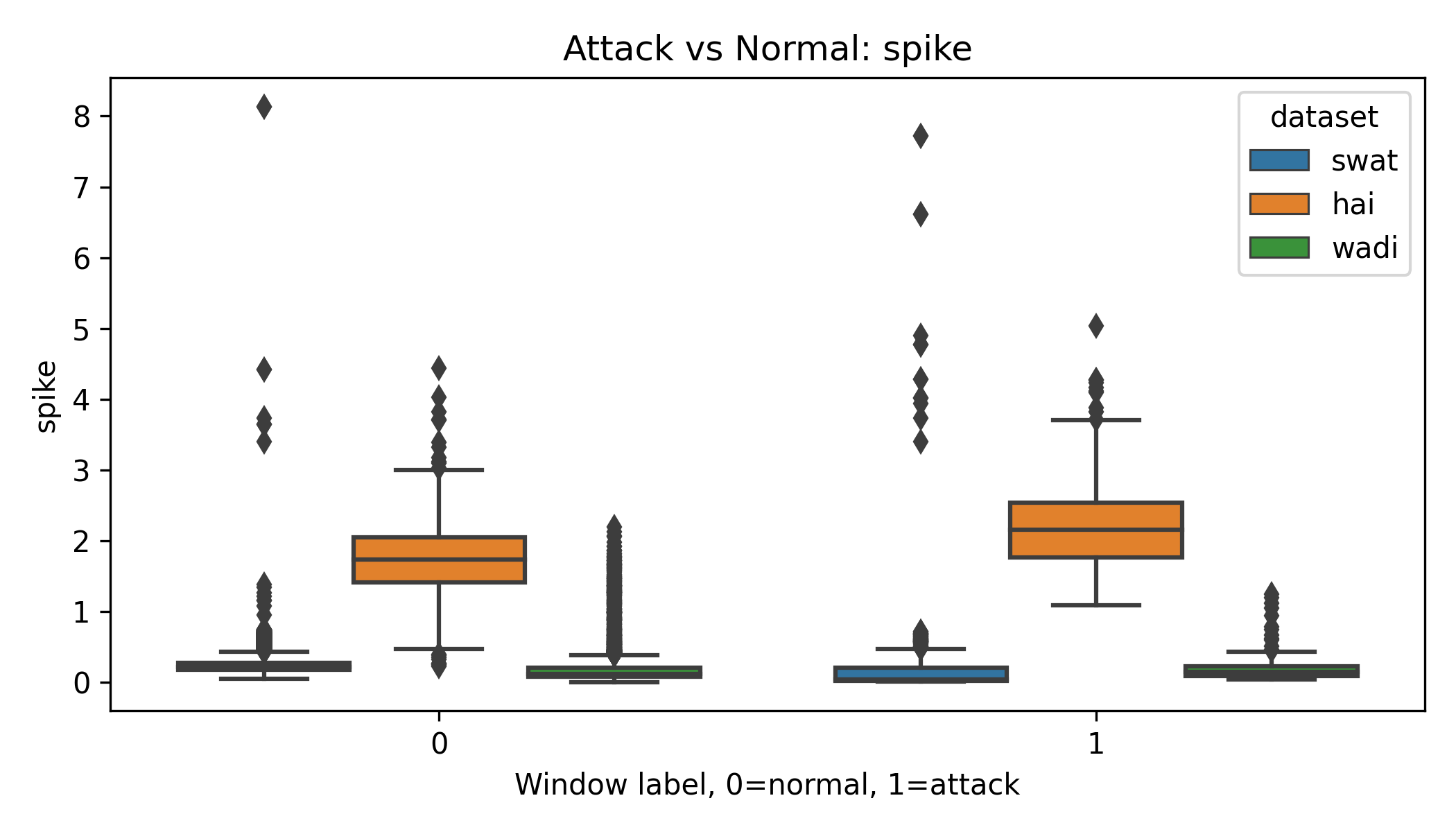}
        \caption{Spike intensity}
        \label{fig:exp1_spike}
    \end{subfigure}
    
    \vspace{0.3cm}
    
    \begin{subfigure}[b]{0.48\linewidth}
        \centering
        \includegraphics[width=\linewidth]{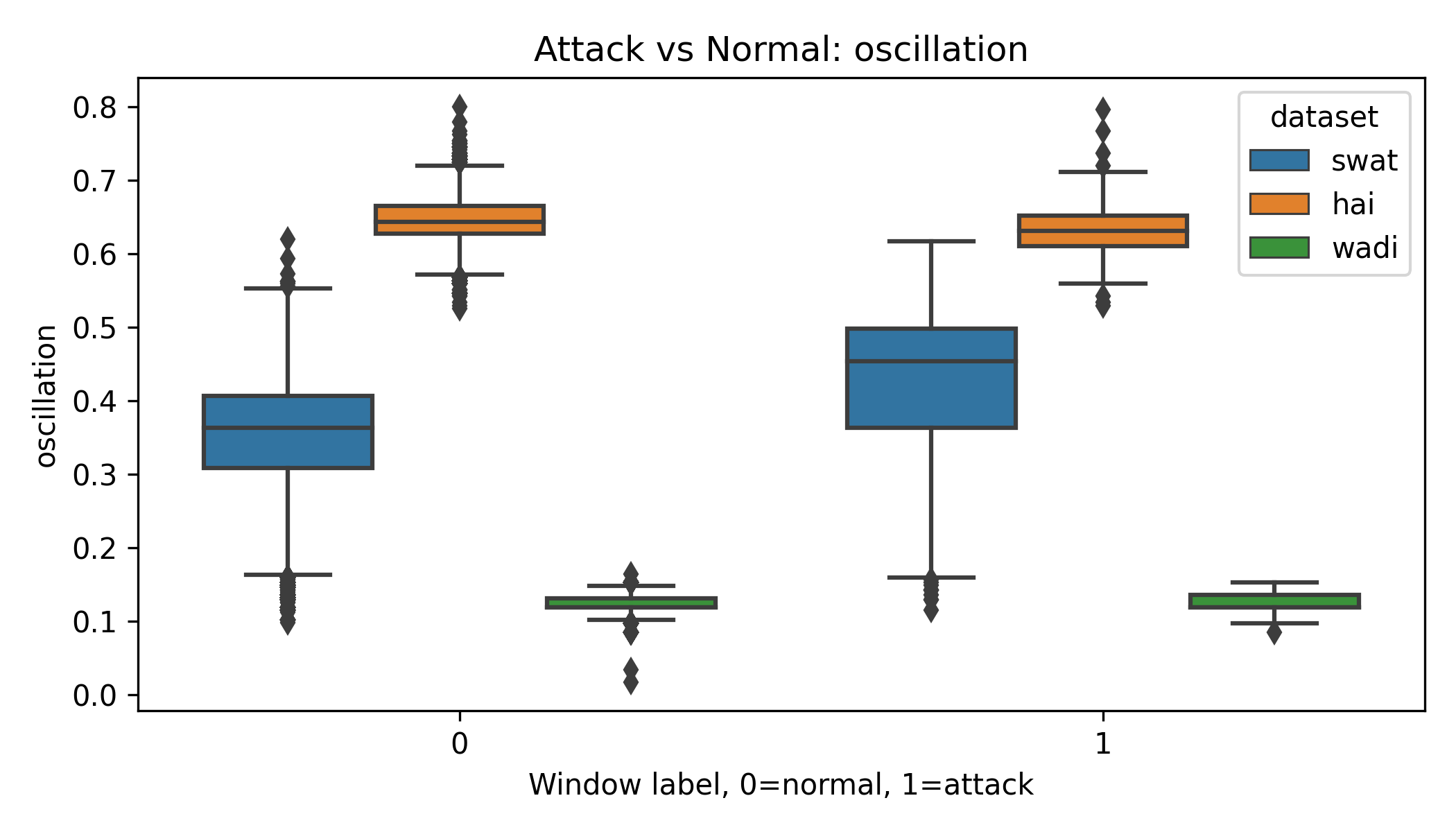}
        \caption{Oscillation intensity}
        \label{fig:exp1_oscillation}
    \end{subfigure}
    \hfill
    \begin{subfigure}[b]{0.48\linewidth}
        \centering
        \includegraphics[width=\linewidth]{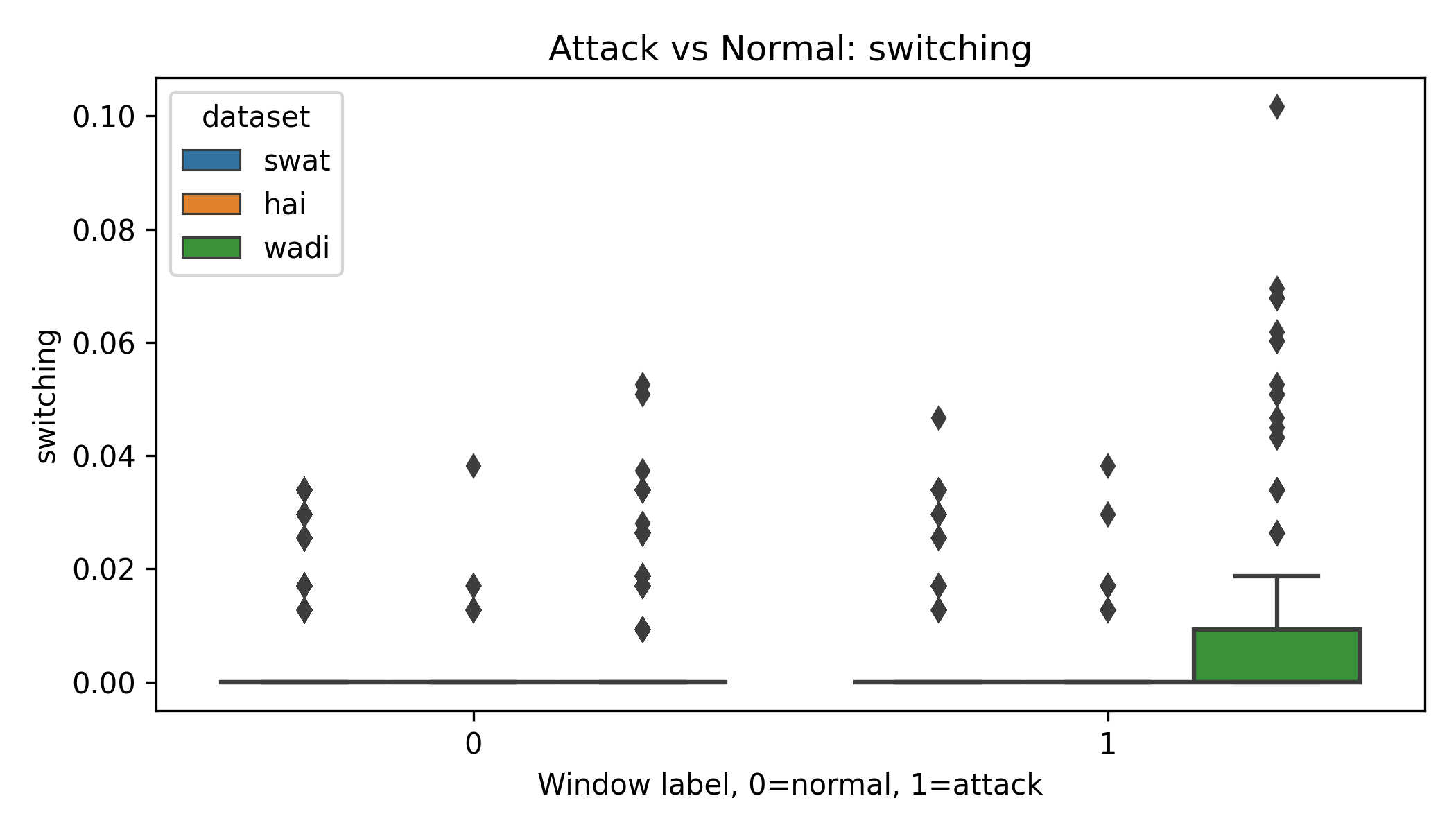}
        \caption{Switching intensity}
        \label{fig:exp1_switching}
    \end{subfigure}
    
    \vspace{0.3cm}
    
    \begin{subfigure}[b]{0.48\linewidth}
        \centering
        \includegraphics[width=\linewidth]{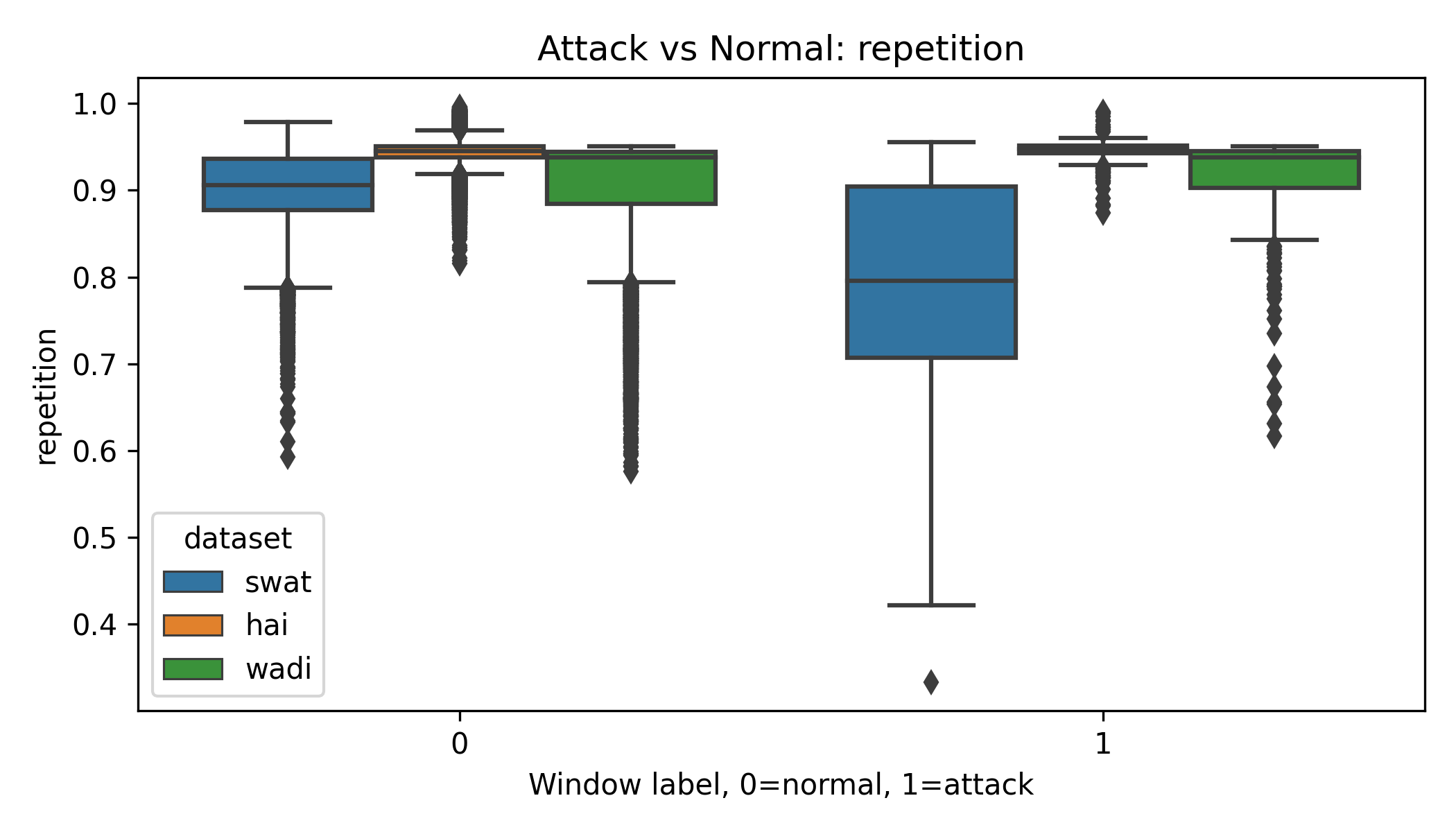}
        \caption{Repetition intensity}
        \label{fig:exp1_repetition}
    \end{subfigure}

    \caption{Distribution of behavioral primitive values for normal (0) and attack (1) windows across the evaluated datasets.}
    \label{fig:attack_vs_normal}
\end{figure*}

To determine whether attack windows exhibit measurable behavioral deviations from normal operation, we first compare the distributions of the proposed behavioral primitives across normal and attack windows for the evaluated datasets, namely SWaT, WADI, and HAI. Figure~\ref{fig:attack_vs_normal} provides a visual overview, while Table~\ref{tab:rq1_statistics} reports median and Interquartile Range (IQR)~\cite{IQR} values together with Mann--Whitney U~\cite{MacFarland2016} test results.

Across the three datasets, attack windows differ measurably from normal windows, but the pattern of change depends on the primitive. As observed in Table~\ref{tab:rq1_statistics}, HAI has the clearest attack-associated increase in both drift and spike, with drift rising from 0.023 to 0.059 and spike from 1.732 to 2.150. SWaT shows a different profile, with oscillation increasing from 0.363 to 0.454 and repetition decreasing from 0.906 to 0.795. WADI exhibits smaller shifts overall, with spike increasing from 0.116 to 0.140, while oscillation remains essentially unchanged at 0.119 in both states. These examples indicate that attack behavior is dataset-dependent rather than uniform across benchmarks.

Collectively, table~\ref{tab:rq1_statistics} confirms that most normal-versus-attack differences are statistically significant, showing also that statistical significance and effect magnitude are not the same thing. For example, WADI oscillation is not significantly different between normal and attack windows ($p=0.732$), and WADI repetition changes only slightly in median terms despite a small $p$-value. Overall, the results indicate that attack windows consistently exhibit behavioral shifts relative to normal operation. Still, the dominant primitive depends on the dataset. Namely, HAI is driven more by spike and drift, SWaT by oscillation and reduced repetition, and WADI by smaller changes centered mainly on spike and drift.

\begin{table*}[t]
\centering
\small
\caption{Distributional comparison between normal and attack windows.}
\label{tab:rq1_statistics}
\begin{tabular}{llccc}
\toprule
Dataset & Primitive & Normal Median [IQR] & Attack Median [IQR] & $p$-value \\
\midrule

SWaT & Drift & 0.003 [0.002, 0.004] & 0.001 [0.000, 0.003] & $<10^{-10}$ \\
SWaT & Spike & 0.216 [0.171, 0.275] & 0.032 [0.019, 0.198] & $<10^{-10}$ \\
SWaT & Oscillation & 0.363 [0.308, 0.407] & 0.454 [0.363, 0.498] & $<10^{-10}$ \\
SWaT & Repetition & 0.906 [0.877, 0.937] & 0.795 [0.707, 0.904] & $<10^{-10}$ \\
SWaT & Switching & 0.000 [0.000, 0.000] & 0.000 [0.000, 0.000] & $<10^{-10}$ \\

\midrule

HAI & Drift & 0.023 [0.014, 0.035] & 0.059 [0.034, 0.174] & $<10^{-10}$ \\
HAI & Spike & 1.732 [1.412, 2.049] & 2.150 [1.761, 2.539] & $<10^{-10}$ \\
HAI & Oscillation & 0.644 [0.627, 0.665] & 0.631 [0.610, 0.653] & $<10^{-10}$ \\
HAI & Repetition & 0.945 [0.938, 0.950] & 0.948 [0.942, 0.952] & $<10^{-6}$ \\
HAI & Switching & 0.000 [0.000, 0.000] & 0.000 [0.000, 0.000] & $<10^{-10}$ \\

\midrule

WADI & Drift & 0.004 [0.002, 0.007] & 0.005 [0.003, 0.008] & $<10^{-10}$ \\
WADI & Spike & 0.116 [0.077, 0.200] & 0.140 [0.085, 0.225] & $<10^{-4}$ \\
WADI & Oscillation & 0.119 [0.119, 0.131] & 0.119 [0.119, 0.136] & 0.732 \\
WADI & Repetition & 0.938 [0.884, 0.945] & 0.938 [0.903, 0.945] & 0.010 \\
WADI & Switching & 0.000 [0.000, 0.000] & 0.000 [0.000, 0.009] & $<10^{-10}$ \\

\bottomrule
\end{tabular}
\end{table*}

\subsection{RQ2: Dataset-Level Differences in Normalized Behavioral Morphology}
\label{SS:rq2}

\begin{figure}[!htbp]
    \centering
    \includegraphics[width=\linewidth]{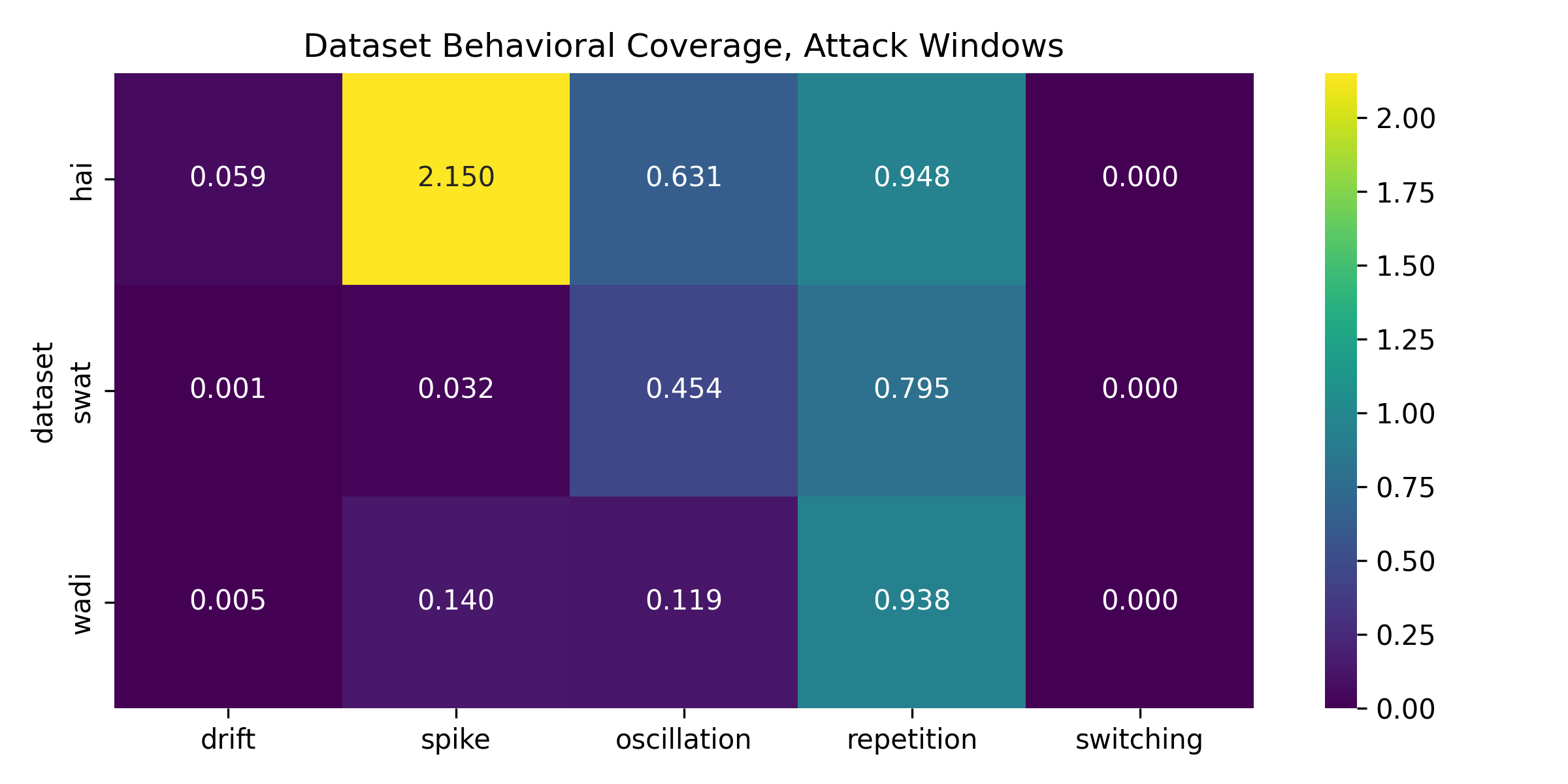}
    \caption{Heatmap of the mean normalized behavioral primitive activations across attack windows, after 95th-percentile aggregation over process variables.}
    \label{fig:heatmap}
\end{figure}

To summarize the attack-window behavior in each dataset, we compute the median activation of each of the five behavioral primitives across all attack windows. Figure~\ref{fig:heatmap} visualizes these summaries, while Table~\ref{tab:rq2_dispersion} reports the corresponding median and IQR values. Recall that the primitives are computed from per-variable normalized signals (Section~\ref{SS:preprocess}) and then aggregated across variables using the 95th percentile (Section~\ref{SS:percentile}); thus, they should be interpreted as unitless descriptions of temporal morphology rather than as directly comparable measures of physical severity across datasets.

Table~\ref{tab:rq2_dispersion} shows that the three datasets occupy different regions of the normalized behavioral space. HAI is the most spike-heavy, with a median spike of 2.150, compared with 0.140 in WADI and 0.032 in SWaT. It also has the highest drift and oscillation values among the three datasets. SWaT places more emphasis on oscillation and repetition, with medians of 0.454 and 0.795, while WADI is dominated by repetition, with a median of 0.938 and comparatively low drift and oscillation. Switching remains near zero in all three datasets. Overall, the table indicates that benchmarking on a single dataset emphasizes a particular normalized behavioral profile and may underrepresent others.

Because the datasets also differ in sampling rate, process characteristics, and temporal granularity, these summaries should not be interpreted as comparisons over equivalent physical time spans. Instead, they describe how each benchmark populates the same normalized behavioral vocabulary. In that sense, table~\ref{tab:rq2_dispersion} provides a compact numerical view of dataset bias in behavioral morphology, rather than a ranking of attack severity across datasets.

\begin{table}[t]
\centering
\caption{Attack-window behavioral coverage summarized as median [IQR] for each dataset and primitive.}
\label{tab:rq2_dispersion}
\begin{tabular}{llc}
\toprule
Dataset & Primitive & Median [IQR] \\
\midrule
SWaT & Drift & 0.001 [0.000--0.003] \\
SWaT & Spike & 0.032 [0.019--0.198] \\
SWaT & Oscillation & 0.454 [0.363--0.498] \\
SWaT & Repetition & 0.795 [0.707--0.904] \\
SWaT & Switching & 0.000 [0.000--0.000] \\
\midrule
HAI & Drift & 0.059 [0.034--0.174] \\
HAI & Spike & 2.150 [1.761--2.539] \\
HAI & Oscillation & 0.631 [0.610--0.653] \\
HAI & Repetition & 0.948 [0.942--0.952] \\
HAI & Switching & 0.000 [0.000--0.000] \\
\midrule
WADI & Drift & 0.005 [0.003--0.008] \\
WADI & Spike & 0.140 [0.086--0.226] \\
WADI & Oscillation & 0.119 [0.119--0.136] \\
WADI & Repetition & 0.938 [0.903--0.945] \\
WADI & Switching & 0.000 [0.000--0.009] \\
\bottomrule
\end{tabular}
\end{table}

\subsection{RQ3: Intra-Dataset Behavioral Diversity}
\label{SS:rq3}

Although Section~\ref{SS:rq2} establishes that the datasets are behaviorally biased away from each other, we further investigate whether attack events are homogeneous within a single dataset. While the heatmap in Figure~\ref{fig:heatmap} captures the dominant average behavioral traits of each benchmark, the presence of non-zero scores across multiple primitives indicates that attacks within the same dataset can induce substantially different temporal process behaviors. This effect is particularly evident in HAI, where attack windows exhibit strong spike, oscillation, and drift activations simultaneously, suggesting the coexistence of multiple behavioral attack patterns. Even in datasets with a more dominant behavioral signature, such as SWaT and WADI, attack windows are not behaviorally uniform and still span multiple temporal primitives. For instance, a single IDS evaluation split may contain both an abrupt transient disturbance and a gradual process deviation. Under standard evaluation protocols, however, both are collapsed into the same binary label (``attack''). The underlying variation in these physical manifestations demonstrates that binary labels may obscure a substantial degree of behavioral diversity, treating a spectrum of attack-induced process behaviors as if they represented a single operational state.

\subsection{RQ4: Behavior-Proxy Separability in Binary IDS Evaluation}
\label{SS:rq4}

Furthermore, to assess how well a baseline classifier separates attack windows according to our normalized behavioral proxies, we first report the standard binary setting and then provide a secondary heuristic multiclass reformulation using the dominant primitive labels defined in Section~\ref{SS:behPriDe}. Note that this multiclass result is descriptive rather than definitive, because it changes the task, the metric, and the target labels simultaneously. Specifically, each attack window was assigned the label of its dominant behavioral primitive, converting the problem from binary to multiclass classification. We emphasize that this is not a new IDS model or an exhaustive tuning exercise; rather, we use an indicative Random Forest classifier as a simple baseline for examining separability under the proposed proxy labels. Accordingly, the results in Table~\ref{tab:rq4_summary} should be interpreted as a descriptive test of within-dataset separability, not as evidence about physical attack taxonomy or operational severity.

\begin{table*}[t]
\centering
\caption{Binary versus behavior-proxy multiclass classification performance across datasets using macro metrics (\%).}
\label{tab:rq4_summary}
\begin{tabular}{lcccccc}
\toprule
& \multicolumn{3}{c}{Binary} &
\multicolumn{3}{c}{Multiclass} \\
\cmidrule(lr){2-4}
\cmidrule(lr){5-7}
Dataset &
Precision &
F1-score &
Recall &
Precision &
F1-score &
Recall\\
\midrule
SWaT    & 96.26 & 85.44 & 79.47 & 47.11 & 37.84 & 36.06\\
HAI     & 84.82 & 76.67 & 71.71 & 58.89 & 58.01 & 62.80\\
WADI    & 94.89 & 85.21 & 77.33 & 48.23 & 42.92 & 38.67\\
\bottomrule
\end{tabular}
\end{table*}

As shown in Table~\ref{tab:rq4_summary}, binary classification yields higher macro scores than the multiclass reformulation in all three datasets. However, as stated before, this decline should be interpreted cautiously because the multiclass setting is a harder prediction problem, uses different averaging over classes, and relies on heuristic labels derived from dominant proxy activations. Therefore, the reduction in macro F1 does not by itself establish hidden failure modes or reveal ground-truth attack semantics. Instead, it indicates that the baseline separates the proposed behavioral proxies unevenly once the task requires distinguishing among them.

The effect is most pronounced in SWaT and WADI. In SWaT, macro F1 decreases from 85.44\% under binary classification to 37.84\% under multiclass classification. In WADI, it decreases from 85.21\% to 42.92\%. HAI shows a smaller reduction, from 76.67\% to 58.01\%, suggesting comparatively stronger separability among the proxy behaviors. Overall, the binary setting shows that the baseline can reasonably well separate attacks from normal windows. In contrast, the multiclass reformulation shows that the same representation does not uniformly separate the behavioral proxies. This makes the multiclass experiment a useful descriptive complement to binary evaluation, but only as a proxy-based separability analysis rather than evidence that binary evaluation hides detection failures.

\section{Key Takeaways}
\label{S:takeaways}

This study introduces a framework for characterizing ICS attack traces using five interpretable temporal primitives: drift, spike, oscillation, repetition, and switching. Applied to SWaT, WADI, and HAI, it shows that attack windows are not a single homogeneous class, but instead exhibit diverse temporal patterns across and within datasets. The framework is intended as a behavioral description, not a taxonomy of attack semantics. A second contribution is the comparison of dataset coverage in this normalized behavioral space. The results suggest that common ICS benchmarks emphasize different temporal morphologies, so evaluation on one dataset may overrepresent some behavioral profiles and underrepresent others. These differences should be interpreted as normalized behavioral shape, not as direct comparisons of physical severity or operational impact. A third contribution is proxy-based behavior stratification as a complement to binary IDS evaluation. The multiclass reformulation is not evidence that binary detection hides failures; rather, it shows how performance varies when the task requires separating heuristic behavioral proxies instead of only attack versus normal. In this sense, the framework adds a lens for studying dataset bias and coverage, rather than replacing standard benchmarking.

\section{Limitations}
\label{S:limitations}

Several limitations should be noted. First, the framework describes observable temporal deviations, but it does not infer attacker intent, semantic attack categories, or causal mechanisms. Second, it uses only L0/L1 process data, so attacks that remain purely cyber and do not affect physical variables cannot be captured. Third, the five primitives were chosen for interpretability and portability, not exhaustiveness. Other descriptors, including spectral, entropy-based, or control-aware features, may capture behaviors outside the current vocabulary. The fixed 60-second window may also smooth short transients and truncate slow degradations. Fourth, the single dominant-label assignment is a coarse heuristic that can hide mixed behaviors when multiple primitives co-occur in the same window. Finally, the primitives are computed on normalized signals and aggregated across variables, so they describe unitless temporal morphology rather than directly comparable physical severity. In other words, similar primitive values may correspond to different operational phenomena across datasets. Hence, cross-dataset comparisons should be interpreted as comparisons of normalized behavioral shape, not physical impact.

\section{Conclusion}
\label{sec:conclusion}

ICS security relies heavily on benchmark datasets to train and evaluate IDS. However, evaluating these benchmarks only through homogeneous binary labels provides an incomplete view of IDS performance, especially when attack traces differ substantially in their temporal manifestations. In this paper, we propose a behavioral characterization framework to examine normalized temporal morphology across SWaT, WADI, and HAI. Our analysis showed that cyber-physical attack windows exhibit diverse process-trace patterns, including abrupt spikes, sustained drifts, sensor freezing, and forced actuator switching, and that standard benchmarks differ in the normalized behavioral profiles they emphasize. These findings indicate that aggregate binary metrics may obscure how performance varies across different proxy-based attack manifestations. At the same time, we do not argue against binary detection itself; rather, we view it as a well-established first stage in a broader escalation pipeline. Our critique is directed at the evaluation practice that treats binary performance as sufficient evidence of robustness, even though behavior-aware triage and diagnosis are rarely instantiated or assessed. 

Accordingly, our framework is meant to complement binary evaluation with behavior-stratified benchmarking, so that variation in normalized temporal morphology is measured and not assumed away. Promising avenues for future work include: i) to expand the behavioral vocabulary with richer descriptors, such as spectral, entropy-based, delay-sensitive, or control-aware features; ii) to move from single-label assignment toward mixed-label or multi-label representations to better reflect co-occurring behaviors and reduce winner-take-all effects; iii) to broader evaluate across more datasets, and ideally, expert-curated annotations to facilitate test generalization and strengthen validation; and iv) to integrate behavioral characterization with IDS triage, alert prioritization, and incident response to make the framework more operationally useful.

\appendix

\section{Sensitivity Analysis}
\label{app:sensitivity}

To assess the sensitivity of the proposed framework to parameter choices, we vary two key parameters: the sliding-window length and the dominant-property threshold. Window lengths of 30, 60, and 120 samples are combined with threshold values $\tau \in \{0.2,0.3,0.4\}$. For each configuration, we report three metrics. First, we quantify the stability of the dominant behavioral labeling across parameter settings by measuring the agreement rate with the selected operating point $(w=60,\tau=0.30)$.

\begin{equation}
A_{\mathrm{dom}}
=
\frac{1}{M}\sum_{j=1}^{M}
\mathbb{I}\!\left[\hat{p}_j = \hat{p}^{\mathrm{base}}_j\right],
\label{eq:dominant:agreement}
\end{equation}

where $\hat{p}_j$ denotes the dominant behavioral property assigned to the $j$-th attack window under a given configuration. $\hat{p}^{\mathrm{base}}_j$ is the corresponding dominant property under the baseline configuration $(w=60,\tau=0.30)$, $M$ is the number of matched windows, and $\mathbb{I}[\cdot]$ is the indicator function, which equals 1 when the two labels match and 0 otherwise. Higher values of $A_{\mathrm{dom}}$ indicate that the same attack window is assigned the same dominant behavioral label across parameter changes, whereas lower values indicate greater sensitivity to the chosen window size or threshold. Second, behavioral diversity is measured via entropy, as shown in equation~\ref{eq:entropy}.

\begin{equation}
H
=
-\sum_{i=1}^{N}
p_i \log(p_i),
\label{eq:entropy}
\end{equation}

where $p_i$ is the proportion of windows whose dominant property is $i$ and $N$ is the number of behavioral categories. Specifically, $H$ summarizes how many primitives are actually exercised, namely, low entropy means that attacks are dominated by one or two behaviors (e.g., mostly spikes), whereas high entropy indicates a richer mix of behaviors (spikes, drifts, oscillations, etc.). Second, we quantify the stability of behavioral diversity relative to the selected operating point $(w=60,\tau=0.30)$.

\begin{equation}
S_H
=
1-
\frac{|H-H_{\mathrm{base}}|}
{\log(N)},
\label{eq:entropy:base}
\end{equation}

where $H_{\mathrm{base}}$ is the entropy at $(w=60,\tau=0.30)$. Values $S_H \approx 1$ indicate that the behavioral diversity remains close to the baseline under parameter changes, while lower values indicate higher sensitivity. As observed in Table~\ref{tab:sensitivity}, entropy remains relatively stable for SWaT and HAI, with only modest changes across window lengths and threshold values. WADI exhibits a stronger dependence on window length, with entropy decreasing for larger windows and increasing for shorter windows, indicating that temporal granularity influences the diversity of dominant behavioral assignments.

In parallel, the agreement rate $A_{\mathrm{dom}}$ is highest at the selected configuration $(w=60,\tau=0.30)$ and remains very close to $1.0$ for nearby settings in SWaT and HAI, confirming that the dominant behavioral label is highly stable in these configurations. WADI also achieves perfect agreement at the baseline configuration and maintains relatively high agreement under larger windows, although agreement decreases more noticeably for shorter windows. Overall, the selected configuration represents a balanced operating point, providing highly stable labeling across all three datasets while minimizing sensitivity to parameter changes.

\begin{table*}[!htbp]
\centering
\caption{Sensitivity analysis for window size and dominant-property threshold. Highlighted rows indicate the parameter configuration used in the experiments of Section~\ref{sec:results}.}
\label{tab:sensitivity}
\begin{tabular}{lc c ccc c ccc c ccc}
\toprule
Window & \(\tau\) & & \(A_{\mathrm{dom}}\) & Entropy & \(S_H\) & & \(A_{\mathrm{dom}}\) & Entropy & \(S_H\) & & \(A_{\mathrm{dom}}\) & Entropy & \(S_H\) \\
\midrule
30  & 0.2 & \cellcolor{lightgray} & 0.808 & 0.509 & 0.903 & \cellcolor{lightgray} & 0.798 & 0.498 & 0.906 & \cellcolor{lightgray} & 0.631 & 0.658 & 0.696 \\
30  & 0.3 & \cellcolor{lightgray} & 0.807 & 0.521 & 0.897 & \cellcolor{lightgray} & 0.796 & 0.512 & 0.898 & \cellcolor{lightgray} & 0.630 & 0.660 & 0.695 \\
30  & 0.4 & \cellcolor{lightgray} & 0.800 & 0.555 & 0.878 & \cellcolor{lightgray} & 0.790 & 0.520 & 0.893 & \cellcolor{lightgray} & 0.625 & 0.665 & 0.689 \\
60  & 0.2 & \cellcolor{lightgray} & 1.000 & 0.336 & 1.000 & \cellcolor{lightgray} & 1.000 & 0.328 & 0.999 & \cellcolor{lightgray} & 1.000 & 0.802 & 0.853 \\
\rowcolor{blue!15}
60  & 0.3 & \cellcolor{lightgray} & 1.000 & 0.336 & 1.000 & \cellcolor{lightgray} & 1.000 & 0.329 & 1.000 & \cellcolor{lightgray} & 1.000 & 0.802 & 0.853 \\
60  & 0.4 & \cellcolor{lightgray} & 0.998 & 0.346 & 0.994 & \cellcolor{lightgray} & 0.991 & 0.329 & 1.000 & \cellcolor{lightgray} & 0.999 & 0.801 & 0.852 \\
120 & 0.2 & \cellcolor{lightgray} & 0.893 & 0.308 & 0.984 & \cellcolor{lightgray} & 0.865 & 0.301 & 0.984 & \cellcolor{lightgray} & 0.808 & 0.468 & 0.797 \\
120 & 0.3 & \cellcolor{lightgray} & 0.893 & 0.308 & 0.984 & \cellcolor{lightgray} & 0.865 & 0.301 & 0.984 & \cellcolor{lightgray} & 0.808 & 0.468 & 0.797 \\
120 & 0.4 & \cellcolor{lightgray}\multirow{-9}{*}{\rotatebox{90}{\textbf{SWaT}}} & 0.893 & 0.314 & 0.988 & \cellcolor{lightgray}\multirow{-9}{*}{\rotatebox{90}{\textbf{HAI}}} & 0.859 & 0.302 & 0.985 & \cellcolor{lightgray}\multirow{-9}{*}{\rotatebox{90}{\textbf{WADI}}} & 0.808 & 0.468 & 0.797 \\
\bottomrule
\end{tabular}
\end{table*}

\begin{credits}
\subsubsection{\ackname}This work is supported by the Research Council of Norway through the SFI Norwegian Centre for Cybersecurity in Critical Sectors (NORCICS) project no. 310105 and by the European Union’s Horizon Europe Research and Innovation Programme through SECASSURED under grant agreement No 101225858 and ENFIELD under grant agreement No 101120657. 
\end{credits}

\bibliographystyle{splncs04}
\bibliography{references}

\end{document}